\documentclass[10pt,conference]{IEEEtran}
\usepackage{epsfig,setspace,amsmath,epsf,amssymb,bm,theorem,cite,graphicx, epstopdf,algorithm,float,color,mathtools, authblk}
\usepackage[table,xcdraw]{xcolor}
\usepackage{subcaption}
\usepackage{algpseudocode}

\newcommand{\RomanNumeralCaps}[1]
    {\MakeUppercase{\romannumeral #1}}
    
\IEEEoverridecommandlockouts
\allowdisplaybreaks
\usepackage{verbatim}

\begin{document}


\title{6G at $\frac{\text{{\Huge 1}}}{\text{{\Huge 6}}} {\bm g}$: The Future of Cislunar Communications \thanks{The authors are with the Department of Electrical and Computer Engineering, at the University of Maryland, College Park, MD, 20742, USA. Emails: \{sahanl, stavros, pmitra, ulukus\}@umd.edu.}}

\author{Sahan Liyanaarachchi \qquad Stavros Mitrolaris \qquad Purbesh Mitra \qquad Sennur Ulukus}

\maketitle

\begin{abstract}
What will the future of cislunar communications be? The ever-expanding horizons of the space exploration missions, and the need for establishing sustainable space communication and navigation infrastructure necessitate to think this question thoroughly. In this article, we examine how some of the concepts of 6G technologies developed for terrestrial networks can be relevant in the context of cislunar networks. We discuss how 6G concepts, such as reconfigurable intelligent surfaces, quantum-resistant physical layer security, private information read/write/cache networks, semantic and goal-oriented communications, information freshness based quality of communication metrics, multi-relay and cooperative networks, hold the potential to shape the future of cislunar communications.
\end{abstract}

\section{Introduction}
Cislunar communication plays a pivotal role in the progress of human civilization towards space exploration via facilitating successful missions and broadening the reach of technology. As the complexity of lunar missions has gradually increased over time, the characteristics of cislunar communications have also progressed accordingly. From early robotic space explorations to crewed Apollo missions to the currently ongoing efforts of building sustainable human bases on the lunar surface, the ever-growing development of missions necessitates robust, reliable, and efficient cislunar communication protocols. 

The term ``cislunar space'' is used to refer to the volume  between the earth and the moon. The average distance between our planet and its natural satellite is 384,400 km, which, to put into perspective, is approximately 40-thousand times the altitude at which passenger planes fly. A signal propagating at the speed of light takes around 1.3 seconds to cover that distance, establishing a lower bound on the communication delay and significantly impacting the design of network protocols, as discussed in Section~\ref{sec:tech}. 

\begin{figure*}
    \centering
    \includegraphics[width=0.95\textwidth]{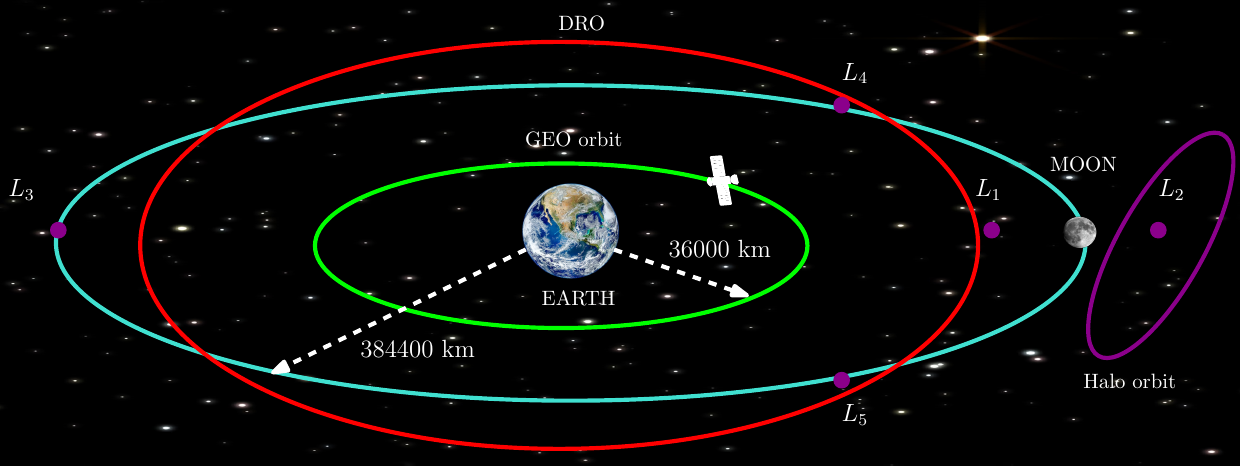}
    \caption{The cislunar space.}
    \label{fig:cislunar_space}
\end{figure*}

In this article, we primarily focus on the earth-orbit-moon communication setting, as shown in Fig.~\ref{fig:cislunar_space} and its slight variants. The cislunar space can be characterized by 5 Lagrangian (libration) points (denoted by $L_1$ to $L_5$) which are unique points in space where the gravitational attraction force from the earth and the moon interact to create an equilibrium. These points remain stationary relative to the earth and the moon, and this enables objects to attain various stable orbits in the cislunar space. One such orbit is the halo orbit around $L_2$ Lagrangian point depicted in purple in Fig.~\ref{fig:cislunar_space}. Distant retrograde orbits (DRO), illustrated in red, are another special class of orbits whose interactions with the $L_1$ and $L_2$ Lagrangian points make them highly stable. DRO objects orbit around earth, however relative to the moon, they form quasi-elliptic orbits that rotate in the direction opposite to the direction the moon orbits the earth. The existence of these stable orbits is instrumental for the deployment of the necessary infrastructure for cislunar communications.

In parallel to the development of cislunar communications, there have been steady progress is terrestrial communications, the most recent version of the communication standard being the fifth generation (5G) communication. The ongoing research work on the next generation, i.e., 6G communications, shows promising aspects of communication that were not considered in previous generations. Examples of such innovative directions include, multi-relay transmission \cite{yuan_towards}, semantic and goal oriented communications \cite{gunduz2023beyond}, freshness and timeliness of information as an emerging quality metric  \cite{yates_age_survey}, reconfigurable intelligent surfaces \cite{koktas2023unleashing}, use of information-theoretic security and privacy measures, such as physical layer security \cite{yener2015wireless, cetin2023secure} and private information read, write, cache \cite{vithana2024private}. In this article, we consider these emerging fields in the context of cislunar communication setting and provide a plausible future direction of how they can be adopted in the relevant future research and development. Non-terrestrial networking (NTN) and use of satellites have already been proposed to supplement terrestrial communication on earth for 6G wireless \cite{halim2018ntn}; multiple layers of various satellite networks around the earth and the moon may be they key to enable seamless 6G and beyond on earth, moon, cislunar space in between, and beyond.

\section{Vision for Cislunar Space: LunaNet}\label{sec:vis}
Historically, lunar missions were sparse and isolated, with significant gaps between successful landings. This trend is observed from the Luna 9 mission in 1966 to the more recent Chandrayaan-3 in 2023. However, the surge in planned missions, particularly those targeting the lunar south pole for its valuable resources, necessitates a robust, scalable, and interoperable communications and navigation (C\&N) infrastructure to support long-term science, exploration, and industrial activities on the moon. As outlined in \cite{reinhart2023lunar}, the future of cislunar infrastructure relies on collaboration between government agencies, commercial, and international partners. 

\begin{figure}
    \centering
    \begin{subfigure}{\columnwidth}
         \centering
         \includegraphics[scale=0.3]{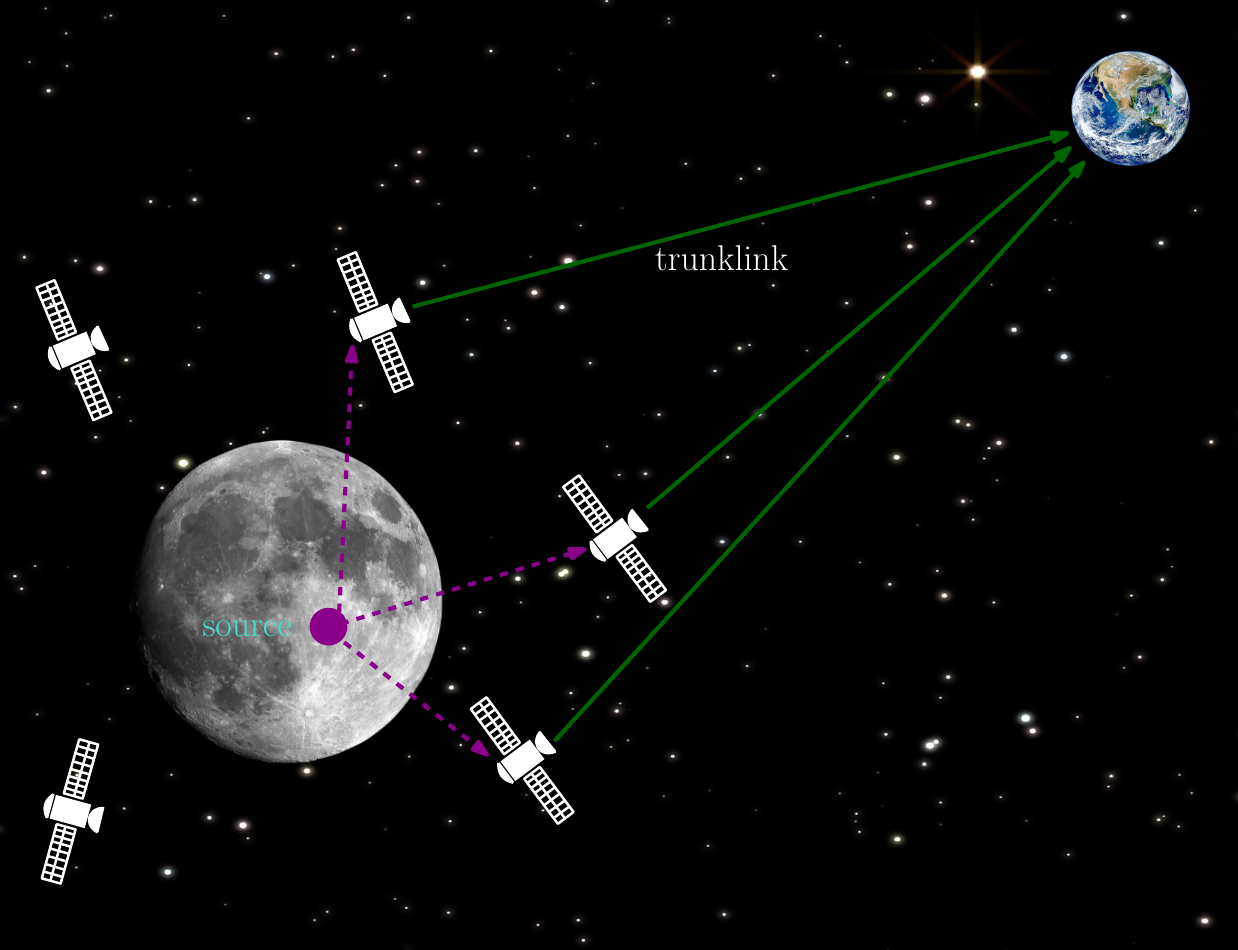}
         \caption{Dedicated trunk link between individual satellites.}
         \label{fig:lunanet_sep}
    \end{subfigure}
    \par\bigskip
    \begin{subfigure}{\columnwidth}
         \centering
         \includegraphics[scale=0.3]{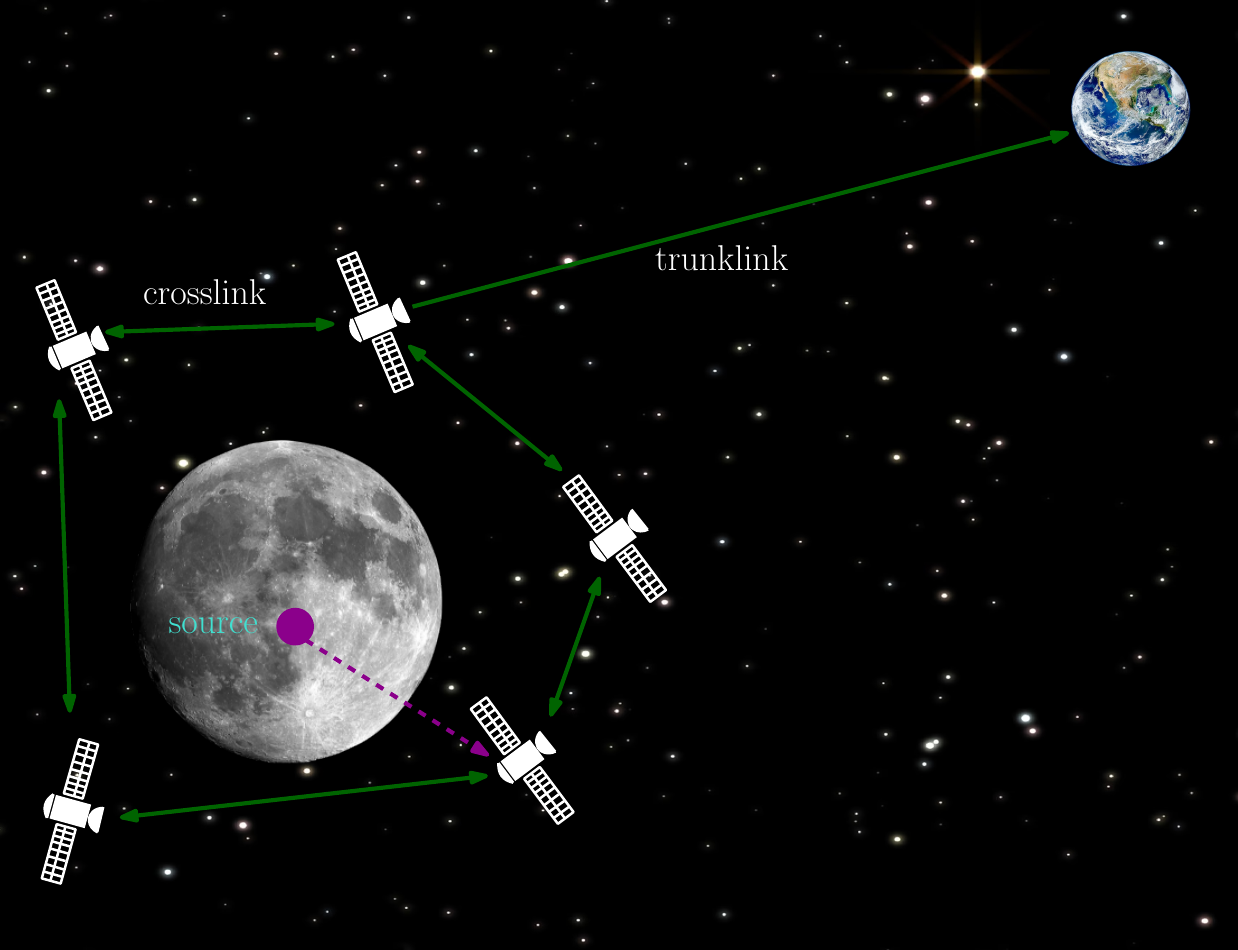}
         \caption{Single trunk link with cross-links between satellites.}
         \label{fig:lunanet_cross}
    \end{subfigure}
    \par\bigskip
    \begin{subfigure}{\columnwidth}
         \centering
         \includegraphics[scale=0.3]{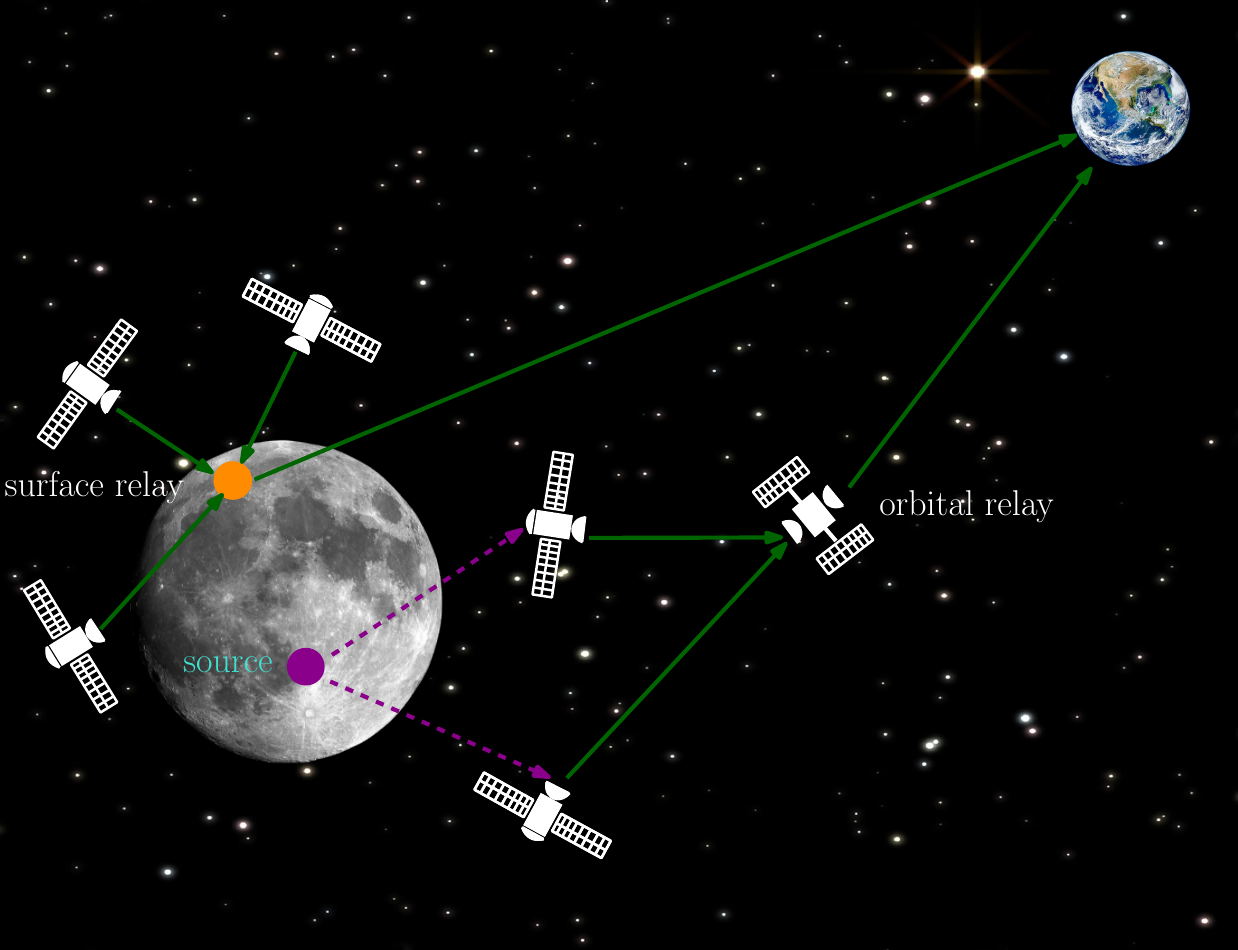}
         \caption{Satellites communicating via a lunar surface/orbital relay.}
         \label{fig:lunanet_surf_orb}
    \end{subfigure}
    \par\bigskip
    \caption{Variations of the  LunaNet node configuration.}
    \label{fig:lunanet}
\end{figure}

The main backbone of this vision is the LunaNet architecture introduced in \cite{israel2020lunanet}. LunaNet is a flexible and extensible C\&N architecture designed to ensure interoperability among various stakeholders. Creating an interoperable global lunar utilization infrastructure is essential, as this system will allow US industry and international partners to maintain a continuous presence on the moon without NASA being the sole user. LunaNet aims to provide key services such as machine-to-machine (M2M) communication, networking, position, navigation, and timing (PNT). The node based architecture of LunaNet enables it to cater to a wide range of applications. Depending on the requirement and constraints of these applications, the nodes of the LunaNet may be arranged in a variety of configurations as shown in Fig.~\ref{fig:lunanet}. The development of LunaNet involves close collaboration with industry partners to refine and implement interoperability specifications, and to ensure that the system can support a diverse range of lunar activities, such as autonomy for lunar robotics, and services for robotic and human missions.

The space communications and navigation (SCaN) program outlines a planned approach to developing a cislunar infrastructure. In the initial stage, the plan is to upgrade the existing deep space network (DSN) antennas and the development of new 18-meter antenna subnets (LEGS) to enhance lunar communication capabilities. This includes increasing data rates and supporting simultaneous operations across multiple frequency bands. Consequently, relay services will be established to ensure continuous coverage, particularly at the lunar south pole and far-side regions. The final goal includes implementation of 3GPP-based cellular technology, optical communications for high bandwidth needs, and the development of lunar navigation services similar to the GPS on earth for the moon.

In \cite{reinhart2023lunar}, the project deliverables are outlined as the following three phases. The \emph{initial phase} (2024-2025) includes establishing foundational PNT communication and navigation services, including direct-to-earth (DTE) and initial relay services. The \emph{growth phase} (2026-2030) includes expanding relay services, introducing optical links, and enhancing surface networking capabilities. Finally, the \emph{desired future state} (2030+) includes achieving a comprehensive and resilient infrastructure with satellite constellations, and extensive surface networks.

Success of the cislunar infrastructure depends heavily on international partnerships and contributions. The interagency operations advisory group (IOAG) and the international communication system interoperability standards (ICSIS) will play crucial roles in coordinating efforts and ensuring that systems developed by different entities can work seamlessly together. 

\section{Emerging Technologies and Concepts}\label{sec:tech}
Humankind has set its sights on establishing a long-term presence on the moon, to facilitate deep space missions and the colonization of mars. This endeavor cannot be successful unless reliable, high-rate communication in the cislunar space is available. With the number of planned moon missions having increased in the last few years, the landscape of communication is expected to be very different from that of the first crewed mission to the moon. Back then, in 1969, a single direct communication link between the earth and the spacecraft sufficed, but in the near future, the cislunar space is expected to host numerous distinct entities, necessitating the development of new communication architectures and technologies. 

\subsection{Optical Communications}
One of the key technologies to enable high-rate communications in the cislunar space is optical communications. Not only do they provide significantly higher data rates compared to radio frequency (RF) communications, but also the corresponding equipment is better suited for space missions. Lasers have a smaller form factor, weigh less, and require less power than RF antennas. NASA has already conducted experiments investigating the strengths and weaknesses of  optical communications and is expected to incorporate an optical communications terminal in the Artemis \RomanNumeralCaps{2} mission, scheduled for after September 2025. The optical downlink will provide rates up to 250 Mbps, allowing base stations on earth to download all the information onboard the spacecraft instead of waiting to obtain the flight recorders from the module returning to the earth. 

\subsection{Lunar Surface Wireless Network}
As more and more lunar assets will be deployed in the future, it is necessary to establish an architecture for lunar surface communications. Relay satellites orbiting the moon could be utilized for this purpose. However, this approach becomes less effective as the number of users grows, and so the need for establishing a lunar wireless network arises. 

Instead of designing separate standardization efforts, this network could be based on existing cellular technologies standardized by the 3GPP, such as 4G/LTE and 5G, which are mature and have proven to be robust \cite{edwards20233gpp}. The aforementioned technologies have been designed for terrestrial users, and adapting them to the space environment is not straightforward; a number of problems need to be addressed. Power consumption in space, unlike on earth, is a principal constraint affecting the design of the equipment. The system also needs to be able to sustain the extreme temperatures and radiation, while the rugged lunar surface with its multiple craters, which need to be explored for valuable resources, can lead to intermittent connectivity. One approach to consider is leveraging the upper layers of the 5G network architecture and tailoring the physical layer for the lunar environment \cite{somerlock2022adapting}.       

In the near future, as part of the Artemis program, Nokia Bell Labs will deploy the first 4G/LTE system at  the moon's south pole \cite{edwards20233gpp}. The system will consist of a 4m-tall base station mounted on a lander, a 1m-tall rover and a free-flying ``hopper,'' utilizing a 20 MHz channel in Band 3 (1710-1785 MHz uplink, 1805-1880 MHz downlink). Even though this will not be a critical system, it will provide invaluable information for future lunar 3GPP-based systems.   

\subsection{Delay Tolerant Networks}
Much like the internet protocol (IP) used in terrestrial communication networks,  an inter-networking protocol known as the delay/disruption tolerant networking (DTN) bundle protocol has emerged to overcome the challenges specific to cislunar space. DTN was introduced in \cite{burleigh2003DTN} to address the issues that arise when using typical internet protocols such as TCP and UDP to transmit data over links with extreme propagation delays. As opposed to the datagrams used in IP, DTN uses data bundles which comprises meta-data such as source-destination identifiers and priority of the bundles. With the use of bundles, DTN employs a \emph{custodial re-transmission} protocol, where  a node  takes custody of a bundle and allocates enough resources to maintain a copy of the bundle until some other node present further in the communication path takes custody of the bundle. Custodial re-transmission along with a non-sequential file transfer protocol known as CFDP enables DTN to provide reliable communication across vast distances. 

\subsection{Reconfigurable  Intelligent Surfaces}
Reconfigurable intelligent surfaces (RISs) have emerged in recent years as a  technology that allows for the construction of intelligent propagation environments. The main idea is that a software-controlled surface steers incident signals towards desired directions, effectively acting as a ``smart'' mirror. This is achieved by partitioning the surface into cells with adjustable properties, which can be altered to change the direction of the reflected signal. In cases where communicating users do not have a clear line-of-sight (LoS) with each other, but they do have a clear LoS with an RIS, a virtual LoS between the users can be established by suitably reflecting incoming signals, consequently, improving the communication quality. Their energy-efficient nature and ease of deployment are additional characteristics that invigorate their potential for use in the lunar surface network. 

A study about the use of RISs on the surface of mars is provided in \cite{koktas2023unleashing} with many of the insights there holding for the lunar surface as well. Strategically positioned RISs can play an important role in mitigating intermittent connectivity by establishing virtual LoS links at a low cost. For instance, in a scenario where a rover explores a crater, communicating with other lunar assets outside will be challenging due to the absence of a clear LoS and the elevation difference. By deploying an RIS on the edge of the crater, the connectivity can be improved in a robust and cost-efficient way since orbiters need not be used. Additionally, RIS can also be utilized when communicating users have a clear LoS but the distance between them is considerable. More sophisticated designs for RISs exist where incident signals are also amplified when reflected, thus, in the interest of preserving energy for mission-critical equipment, such variants can be advantageous. Utilizing RISs to enhance the performance of the localization system in scenarios where lunar assets do not have a LoS with orbiting satellites is another use case highlighted in \cite{koktas2023unleashing}. The development of suitable algorithms constitutes a promising research avenue.

\subsection{New Metrics: Semantics, Goal/Task-Oriented, Timeliness}
Another interesting development considered for 6G, which can be impactful in cislunar communications as well, is the set of emerging novel metrics, such as semantics of communication, goal-oriented and task-oriented communications \cite{gunduz2023beyond}, as well as timelines, freshness and age of information \cite{yates_age_survey}. Traditionally communication networks have been designed with the mindset of transmitting bits from one point to another; and keeping either a high throughput or a low delay in this transmission. In emerging 6G, as well as in cislunar communications, rather than transmitting bits from one point to another, accomplishing a goal could be a more suitable metric. With robots, machines, sensors, humans collaborating in missions, semantics/context/meaning of transmitted information and the purpose and end-to-end completion of the mission (goal-, task-oriented communication) will be critical. As importantly, going beyond the traditional metrics of throughput and delay to more mission-relevant metrics such as age and freshness of information will play important roles.

\section{Multi-Relay Communication} \label{multi-relay}
In addition to the delay tolerant networking described in Section~\ref{sec:tech}, adequate cislunar infrastructure must be integrated into the cislunar communication system to provide enough coverage  even to lunar  surface probes and sensors stationed within the outskirts of the lunar far-side (LFS). Due to a natural phenomenon known as tidal locking, LFS does not have a direct visible link with the earth. Moreover, due to the large delays present in cislunar communications, transmitting data back to earth in a timely manner is crucial for time sensitive scenarios such as crewed lunar landings on the LFS. The use of a single surface relay or a single orbiting relay may not be sufficient to provide coverage to all the equipment present on the lunar surface. Further, all these lunar surface probes and relays work in a severely energy-constrained environment, which limits their rate of data transmission. Thus, the design of multi-relay satellite constellations and scheduling policies to achieve timely communication in an energy-constrained environment arises as a fundamental problem when designing the cislunar infrastructure.

\subsection{Constellation Design}
The satellite constellations must be designed to maximize the lunar coverage while positioning them in minimal energy orbits (i.e., orbits that require the minimal number of manoeuvres to stay in orbit) which in turn minimizes the fuel consumption and maximizes the mission duration of the satellites. Due to the limited orbital stability of conventional lunar orbits, the design of halo orbit constellations have been widely studied in the literature due to the stationarity of the Lagrangian points and the large surface visibility provided by the halo orbits. Halo orbits with higher out of plane amplitude have high orbital stability, however, they may decrease the lunar equatorial coverage. On the contrary, DROs which also exhibit stable orbits, provide sufficient coverage on the lunar equator but lack the coverage at poles. Combining the complementary strengths of the two orbits, \cite{lunar_coverage2020} introduces the design of hybrid halo-DRO satellite constellations, where they show that a four halo orbit satellite constellation or a four DRO satellite constellation can provide total lunar surface coverage. There have been promising research in this direction, where the design of multiple orbit constellations, elliptical inclined orbit constellations and lunar frozen orbit relays are explored.

\subsection{Scheduling Transmissions}
In many remote monitoring systems, if the data from  sensors are not communicated to the remote monitor in a timely manner, the estimates of the parameters maintained at the remote monitor tend to drift away from the actual values. The greater the delay experienced, the greater will be the estimation error. Due to the significant delays already present in the cislunar space, it is clear that timely communication needs to be a vital feature in cislunar communication systems. Moreover, in contrast to its terrestrial counterparts, cislunar communication systems operate under energy-critical environments which limits the rate at which they can communicate. Therefore, the sources must be sampled and transmitted to maximize the timeliness of data while subjected to a sampling rate constraint and an energy budget. 

Age of information (AoI), has been used in a variety of terrestrial communication systems, as a metric quantifying timeliness of communication. AoI measures the staleness of the update at the receiver by considering the time that has elapsed from the time of its generation \cite{yates_age_survey}. Therefore, time averaged AoI stands out as a promising candidate for evaluating the design of network architectures and the performance of the sampling policies to be used in cislunar space.

Most delay-based communication systems are often modelled as queuing models, where the service time of the server accounts for the delay experienced. Much of the established literature studies the design of sampling policies for a single source single server queue to minimize the average AoI. However, recent developments have shown that the use of multiple servers significantly improve the AoI performance of the system. This is one of the underlying motives behind the development of multi-relay communication networks in cislunar space. However, due to limited infrastructure present in the cislunar space, these multi-relay networks are often shared by multiple sources. When a single relay is shared by multiple sources, the source transmissions must be scheduled appropriately. Therefore, the sampling policies must be designed in coherence with the scheduling policies.

\begin{figure}
    \centering
    \includegraphics[width=0.8\columnwidth]{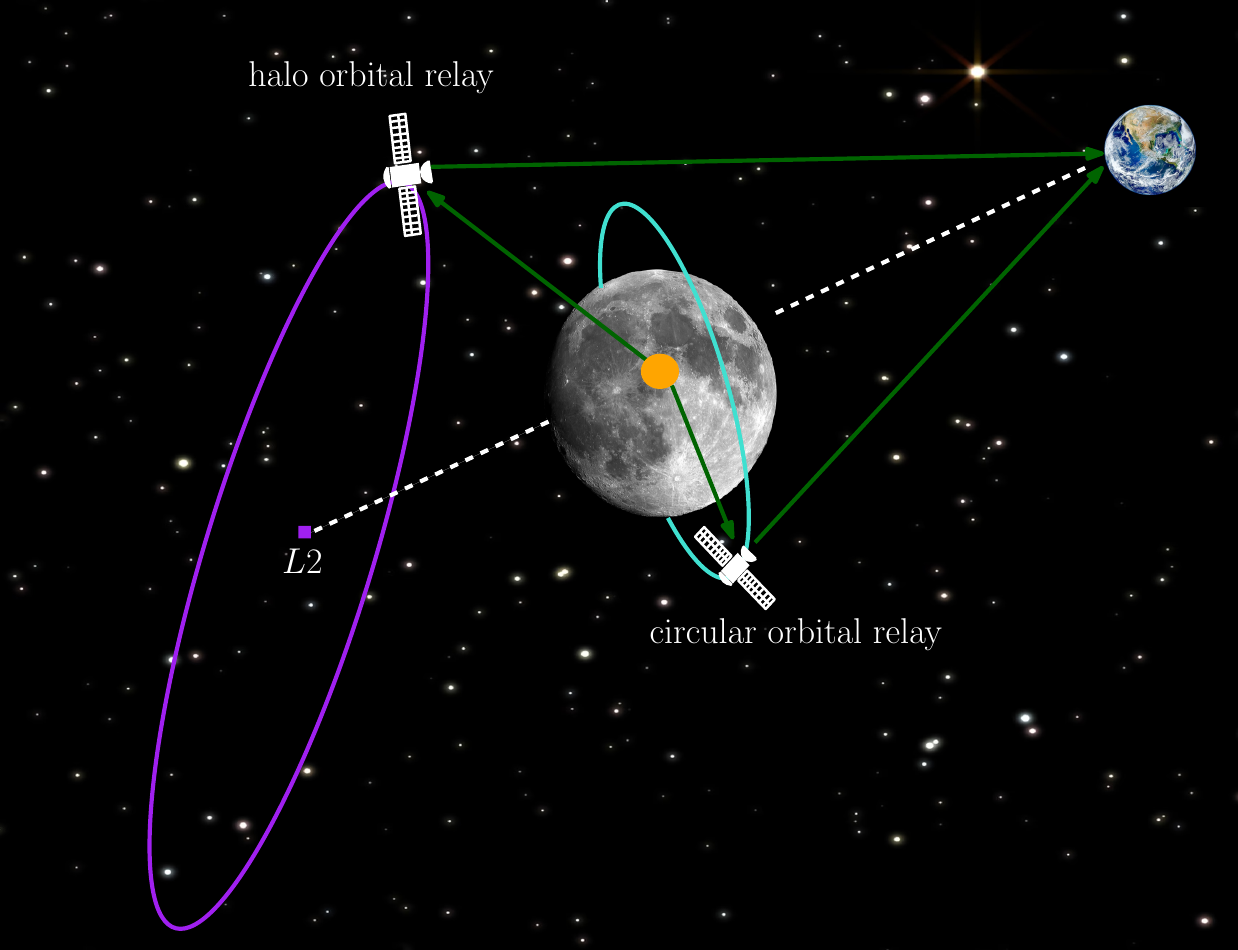}
    \caption{Hybrid halo-circular orbital satellite constellation.}
    \label{fig:cir_halo}
\end{figure}

The work in \cite{yuan_towards} explores the above research direction, by using the AoI metric for the design of a satellite constellation to communicate with the LFS. They propose a framework for the design of a hybrid constellation which utilizes both a circular orbital satellite and a halo orbital satellite placed at the Lagrange L2 point as shown in Fig.~\ref{fig:cir_halo}. The use of a halo orbital satellite provides the coverage to LFS, which does not have direct visibility to the earth, while enabling sensors within the coverage of both the satellites to relay their data through multiple paths. By considering this to be a multi-source two-server queuing model, they design policies for scheduling the source transmissions between the two satellites.

Drawing inspiration from the terrestrial counterparts, we present the following variants of the queuing models, that may be useful to characterize some of the natural attributes present in cislunar communication systems:

\begin{itemize}
    \item We note that for some sensors (e.g., sensors present in the lunar equatorial region), the circular orbital satellites may only be available periodically. As opposed to fixed base stations or relays present in most terrestrial communication systems, in cislunar space, since most relays are orbital satellites, these types of periodic unavailability periods are more common. These types of relays can be modelled using servers that take vacations periodically.
    \item Some satellites may be equipped with both optical and RF transmitters. The former provides significantly higher data rates but is less robust as its operation depends on the weather conditions on earth. This system, with its discrepancy in service times, can be modeled as a queue with a backup server.
    \item Another important criteria that must be addressed is the effect of energy availability on the transmission rate. The main source of energy for the lunar probes and relays comes in the form of solar energy. During lunar nights (periods of time not exposed to the sun), the sensors and relays operate with a low energy budget, and hence, the rate of transmission must be reduced. This can be modelled by servers whose service rate is a time-dependent function, while the sensors can be modelled as energy-harvesting nodes.
\end{itemize}

Even though AoI is a good indicator of the timeliness of the communication system, we note that, when possible, the age of incorrect information (AoII) is a more suitable metric for the design of scheduling policies for cislunar space. AoII metric incurs a penalty based on the time duration at which there is a discrepancy between the source and the destination. Therefore,  AoII-optimal scheduling policies would try to transmit the data only if there is a mismatch between the estimates, as opposed to AoI-optimal policies which try to transmit most current information (without considering its importance) to the destination. This enables AoII-optimal policies to operate at lower transmission rates, and therefore, would be useful in energy-constrained environments such as the cislunar space.

The design of age-agnostic scheduling policies would also be pivotal for the cislunar space, since they do not require an explicit feedback channel. Age-dependent policies on the other hand would require an exact replica of the associated age process, which can be difficult to maintain due to the long propagation delays and packet errors, and therefore, should almost always be supplemented with an ARQ-like protocol. Hence, we  believe the design of age-agnostic policies such as probabilistic schedulers and cyclic schedulers would play an important role in the future of cislunar communications. 

\begin{figure}
    \centering
    \includegraphics[width=0.8\columnwidth]{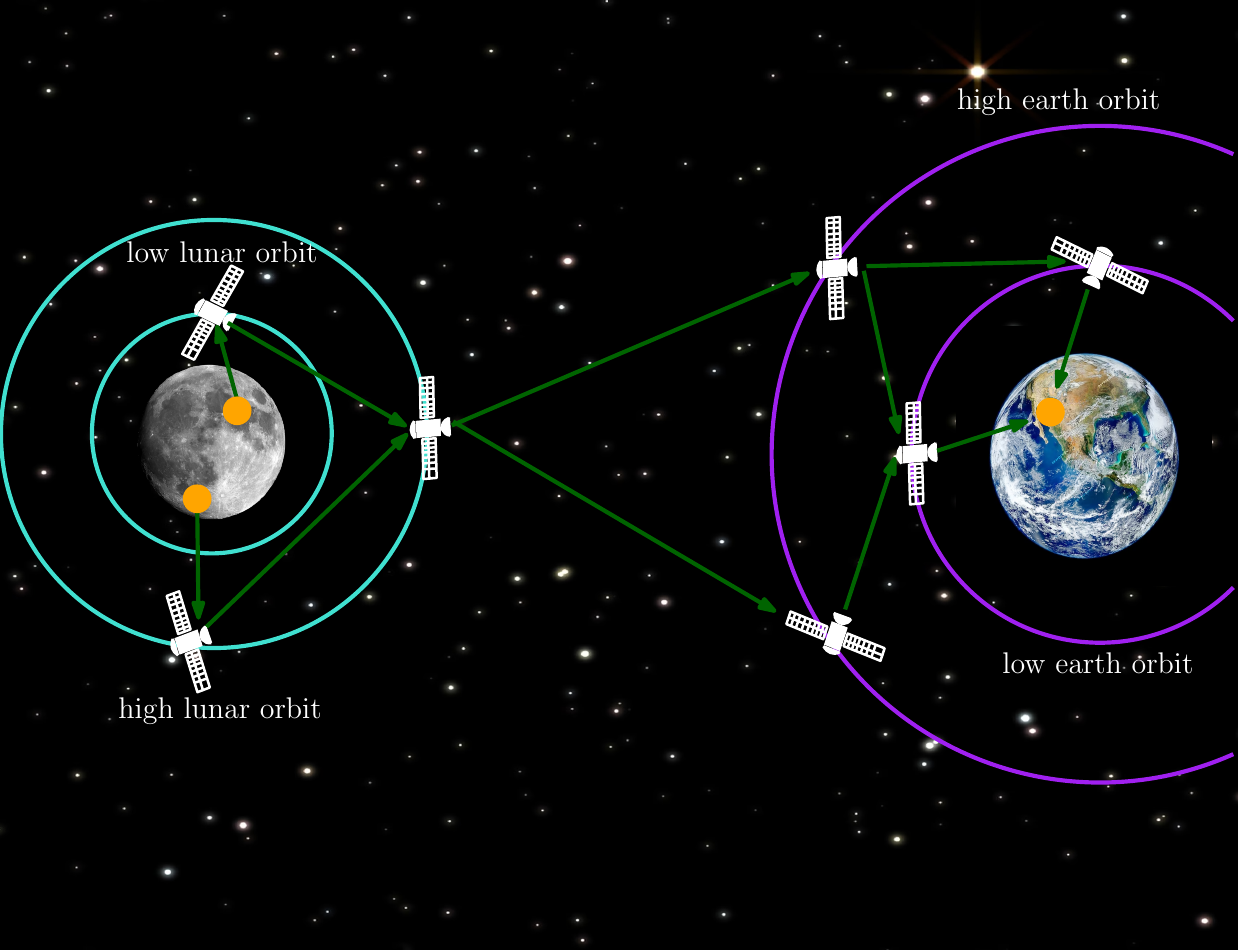}
    \caption{Multi-hop satellite relay network.}
    \label{fig:multi_hop}
\end{figure}

\subsection{Multi-Hop Relays}
Due to the long propagation distances, the signal strength of the transmitted data drastically decreases the further it is from the source. This along with solar interference can heavily degrade the signal-to-noise ratio (SNR) of the signals at the earth observatories. To mitigate this issue, an ideal choice is to relay the signal via multiple hops. To accomplish this, satellite orbits must be established at multiple altitudes both around the earth as well as around the moon. Existing low-earth orbit and high-earth orbit satellites can be refurbished, and new lunar orbital satellites must be established for this purpose. However, the establishment of new orbital satellite must be carefully chosen, since the gravitational influence of the sun and jupiter may disrupt the stability of their orbits. Fig.~\ref{fig:multi_hop} envisions one such multi-hop relay network.

\subsection{Caching}
Another technique that can greatly improve the timeliness of communication is the use of caches where more frequently accessed data can be stored in intermediate relays to minimize the latency. From an information freshness perspective, cached data is advantageous in minimizing transmission delays. However, if the cached data becomes out-of-date, it can adversely affect the freshness of the information received. Thus, outdated data must be evicted and replaced with fresh data. Focused on this phenomena, caching have been extensively studied in information dissemination networks where the optimal cache update rates and cache update policies have been developed to maximize information freshness. Thus, these caching algorithms emanating from information dissemination networks would be instrumental for the establishment and proper placement of orbital caches (caching relays) in the cislunar space. We envision, that these orbital caches would be the gateway between the earth-moon communication links in the future.

Depending on the altitudes of these caching relays, they may be so expensive that no single company or a single country may be able to afford their launch, or even if they do, they may not be financially sustainable in the long-run. Thus, many of these relays may have to be shared among different entities (e.g., different corporations or different countries). Consequently, necessary steps must be taken so as to maintain data privacy for these multi-tenant caches. In this regard, information-theory based techniques such as private-read-update-write (PRUW) and private information retrieval (PIR) could be used such that even though multiple entities share the cache or have access to the cache, privacy and integrity of cached data may be guaranteed.

\section{Secure Communication and Storage}
Different entities in the cislunar space are expected to cooperate and form networks, as discussed in Section~\ref{sec:vis} and Section~\ref{multi-relay}, with secure communications being the backbone of this synergy. The subject of security is rather broad but in this section we view it through the lens of wireless communications and information theory. 

\subsection{Interference Detection in Space}
The broadcast nature of RF signals renders cislunar communication systems susceptible to malicious interference attacks, such as jamming and spoofing. 
In addition to disrupting communication, these attacks lead to the wastage of scarce resources, such as energy. Both classical and machine learning (ML)-based techniques have been developed for RF interference (RFI) detection in terrestrial wireless networks, but the physical phenomena in space, which are absent in classical settings, add a new dimension to the problem, necessitating the reevaluation and adaptation of existing methods. 

Towards that end, a limited number of works exist. In \cite{cetin2023secure}, the ML methods proposed for RFI detection use synthetic data that take into account some of the aforementioned physical phenomena, such as the effect of the brightness temperature of the lunar surface on the signal noise.  A proof of concept that ML methods for RFI detection can be effectively deployed in the resource-constrained environment of space is provided in \cite{lefcourt2022space}, paving the way for further research on the topic.

\subsection{Addresing the Quantum Threat}
Currently, secure transmission of information relies on cryptographic schemes. Their success is based on the premise that adversaries lack the computational capabilities required to solve the complex mathematical problems (such as factoring and discrete logarithms) underlying encryption. Despite the tremendous success of these schemes, the perpetual development of quantum computers renders them incompatible with the vision for long-term secure space communications. As we embark on a new era of space exploration, it is worth considering the adoption of information-theoretic approaches that ensure security, irrespective of the computational capabilities of adversaries, by incorporating the requirement for secure communication in the design of the physical layer. 

The main idea behind physical layer security is to guarantee that the observations obtained by an eavesdropper and the secret message of the transmitter are statistically independent; with this constraint in place, computational power becomes irrelevant. A large body of research on the topic exists, with results that are of interest in cislunar communications. 

As discussed in \cite{yener2015wireless}, multiple transmitters can enhance the rate of secure communication by means of cooperative jamming and structured signaling. Even more importantly, these ideas extend to scenarios that include untrusted relay nodes. Given the high cost of deploying relay satellites in lunar orbit, adopting such technologies for secure communications, where relays are treated as untrusted nodes, can facilitate the cooperation between different space agencies and companies. 

While promising, existing results come at the expense of reduced transmission rates, which can be improved by use of multiple antennas or RISs; and are mostly information-theoretic in nature with some polar-code-based implementations, which can be made more practical with design of explicit codes. It is crucial to note that successfully developing physical layer secure space communication systems can revolutionize terrestrial systems. That is, lessons learned in space, where stakes are high, can facilitate the development of analogous terrestrial technologies.
  
\subsection{Private Information Read, Write, Cache}
Another technology that can facilitate the cooperation of different parties in space missions is private information retrieval (PIR) and its extensions, such as, private read-update-write (PRUW), private storage, private compute, private search, private learn, that is, private multi-party computation, in general. As the term suggests, PIR allows users to download data entries from a shared system of databases without the databases being able to determine which entries were actually desired. This is achieved by sending carefully crafted queries to the different databases. The individual answers from these databases do not leak information, but when combined and processed by the user, they yield the desired data. 

Such approaches are easy to implement on the application layer and can be extremely useful in scenarios where different parties share valuable equipment but do not fully trust each other. An important extension of PIR is PRUW which allows users to download, update, and upload the updates back to a storage system without revealing the content downloaded, uploaded or their positions. These features are especially beneficial for securely updating cache entries in shared satellites without revealing the details of the updates.

The main obstacle in adopting these technologies is the requirement of having multiple databases with a set of common entries that the user can query. Given the limited assets in space, storing data across multiple databases may currently be a challenge. However, as our lunar and cislunar infrastructure improves, these methods could become prime candidates for enhancing security and trust-free cooperation. This also points to the need to further study single-database PIR.

\subsection{Security and Scheduling}
In cislunar space, the aspect of secure communications can significantly impact scheduling polices.
By monitoring transmission patterns of relay satellites, adversaries can obtain valuable information. For instance, unusually long periods of communication may indicate significant events, such as emergencies. By leveraging that, adversaries can choose when to launch an attack so as to increase the incurred damage. Given the critical role of relay satellites, their periodic coverage of areas and their limited resources, developing transmission and caching policies that also take into account security is an interesting research direction. 

\section{Conclusion}
We presented existing standards of communications and discussed key challenges specific to cislunar communications. We drew relations from existing terrestrial communication technologies and articulated their correspondence in establishing the next generation of communications for cislunar space. We introduced new research directions that take the intricate aspects of cislunar communications into account. We hope that the research problems listed could be a catalyzer for further research in this area and contribute towards the future of cislunar communications.

\bibliographystyle{unsrt}
\bibliography{refs-used-short}

\end{document}